\documentclass[aps,prb,twocolumn,floatfix,amsmath,amssymb]{revtex4}
\usepackage[final]{graphicx}
\usepackage{bm}
\usepackage{float}
\usepackage{afterpage}

\newcommand{\bk}{{\bf k}}

\newcommand{\br}{{\bf r}}

\newcommand{\ba}{{\bf a}}

\newcommand{\BGamma}{{\bm \Gamma}}

\begin{document}
\title{Topological Insulators on the Lieb and Perovskite Lattices}
\author{C. Weeks}
\affiliation{Department of Physics and Astronomy, University of British Columbia, Vancouver, BC, Canada V6T 1Z1}
\author{M. Franz}
\affiliation{Department of Physics and Astronomy, University of British 
Columbia, Vancouver, BC, Canada V6T 1Z1}
\begin{abstract}
Electrons hopping on the sites of a two-dimensional Lieb 
lattice and three-dimensional edge centered cubic (perovskite) lattice
are shown to form topologically non-trivial insulating phases
when spin-orbit coupling is introduced. These simple models on lattices with cubic symmetry show a Dirac-like structure in the excitation spectrum but with the unusual
feature that there is a dispersionless band through the center of the spectrum and only a 
single Dirac cone per Brillouin zone. \end{abstract}

\maketitle
\section{Introduction}
The study of topological insulators\cite{moore1,hasan1} has taken off in the last few years, proving to be an exciting area of research in condensed 
matter physics, while also having ties to fundamental physics due to the possibility of providing a testbed for exotic particles like Majorana fermions, \cite{Fu3} fractionally charged vortices,\cite{seradjeh1} axions \cite{Qi2,Essin} and magnetic monopoles.\cite{Qi, Franz3}

 In general, topological insulators are bulk insulators having an energy gap between
the valence and conduction bands, but with gapless
edge states (2D) or surface states (3D), caused by spin-orbit coupling, which are immune to non-magnetic impurities and geometric perturbations. Similar in spirit to the quantum Hall effect, the novelty of a 
topological insulator (TI) lies in the
fact that it can be characterized by a topological invariant, and is not the result of a spontaneously broken symmetry.   However,  the invariant can not take on any integer value, as in the in the quantum Hall case, but is instead 
a so called $Z_2$ topological quantum number that we label $\nu$, which can be either 0 or 1. In 2D a single $Z_2$ invariant is sufficient for the job, whereas in 3D one needs 4 $Z_2$ invariants $(\nu_0;\nu_1\nu_2\nu_3)$, which distinguish 16 possible topological phases and fall into the two general classes: weak (WTI) and strong (STI) topological insulators.\cite{Fu} In 3D if there are an odd number of surface states then we have $\nu_0=1$ and the system is in the STI phase. If there is an even number of surface states (possibly zero)  then $\nu_0=0$ and we have a trivial insulator or WTI if any of the $\nu_i$ are nonzero. 

The 2D topological insulator, also known as the quantum spin Hall (QSH) phase, was independently proposed by two groups,\cite{KM1,KM2,Zhang1} and to date a variety of models have been established that can support  the TI phase. The original model on the honeycomb lattice was realized by superposing two copies of the  Haldane model,\cite{Haldane} one for spin up and 
other for spin down electrons, with the
spin up and down electrons moving in opposite directions along the edge.\cite{KM1} It was then predicted theoretically,\cite{Zhang2} and later confirmed experimentally,\cite{Konig} that this effect is present in the HgTe/CdTe quantum wells. Fu and Kane extended the concept to 3D with a toy model on the diamond lattice,\cite{Fu} and also predicted that the alloy Bi$_{1-x}$Sb$_{x}$, which has large spin orbit coupling,  would be a three dimensional TI.\cite{Fu2} This was later verified experimentally by ARPES measurements which directly observed the topological surface states.\cite{Hasan} Other models with a 
low energy Dirac structure, namely the 2D kagome and 3D pyrochlore systems, have been recently shown to support topological states. \cite{Franz1,Franz2,pesin1}
Experimentally the most promising system is currently Bi$_{2}$Se$_{3}$, which is known to have a large bandgap and a single surface Dirac cone. \cite{HasanII}. Although not experimentally
verified, there is also some theoretical work on topological insulators in cold atoms, which looks promising.\cite{DasSarma, Zhang3}

On a practical level, these gapless surface states are robust against disorder, since time reversal symmetry disallows back scattering from non
magnetic impurities. When gapped by time-reversal symmetry breaking perturbations, superconducting or excitonic pairing, these surface states give rise to interesting insulating phases with exotic quasiparticle excitations already mentioned above. Electric manipulation of the spin degrees of freedom
in topological insulators should also be possible and they can therefore be of value in spintronics applications. Some of the existing topological insulators are also known to
be among the best thermoelectric materials, such as the material Bi$_{2}$Te$_{3}$ discussed by Zhang et. al., \cite{Zhang4} and thus discovery of new materials or further understanding of existing ones would be of value.

In this paper, we present another possible host for a topological insulator, namely the Lieb lattice in 2D and its 3D counterpart the perovskite or
edge centered cubic lattice. We show that a simple tight-binding model for electrons on these lattices leads to TI behavior in the presence of spin-orbit coupling. These toy models are novel in that unlike most of the existing models, they have a simple cubic symmetry, and exhibit 
only a single Dirac node in the spectrum intersecting a degenerate flat band precisely at the Dirac point. We also briefly discuss other perturbations that lead to topologically trivial gapped phases.

\section{Lieb Lattice}
\begin{figure}[t]
\begin{center}
\includegraphics[scale=0.198]{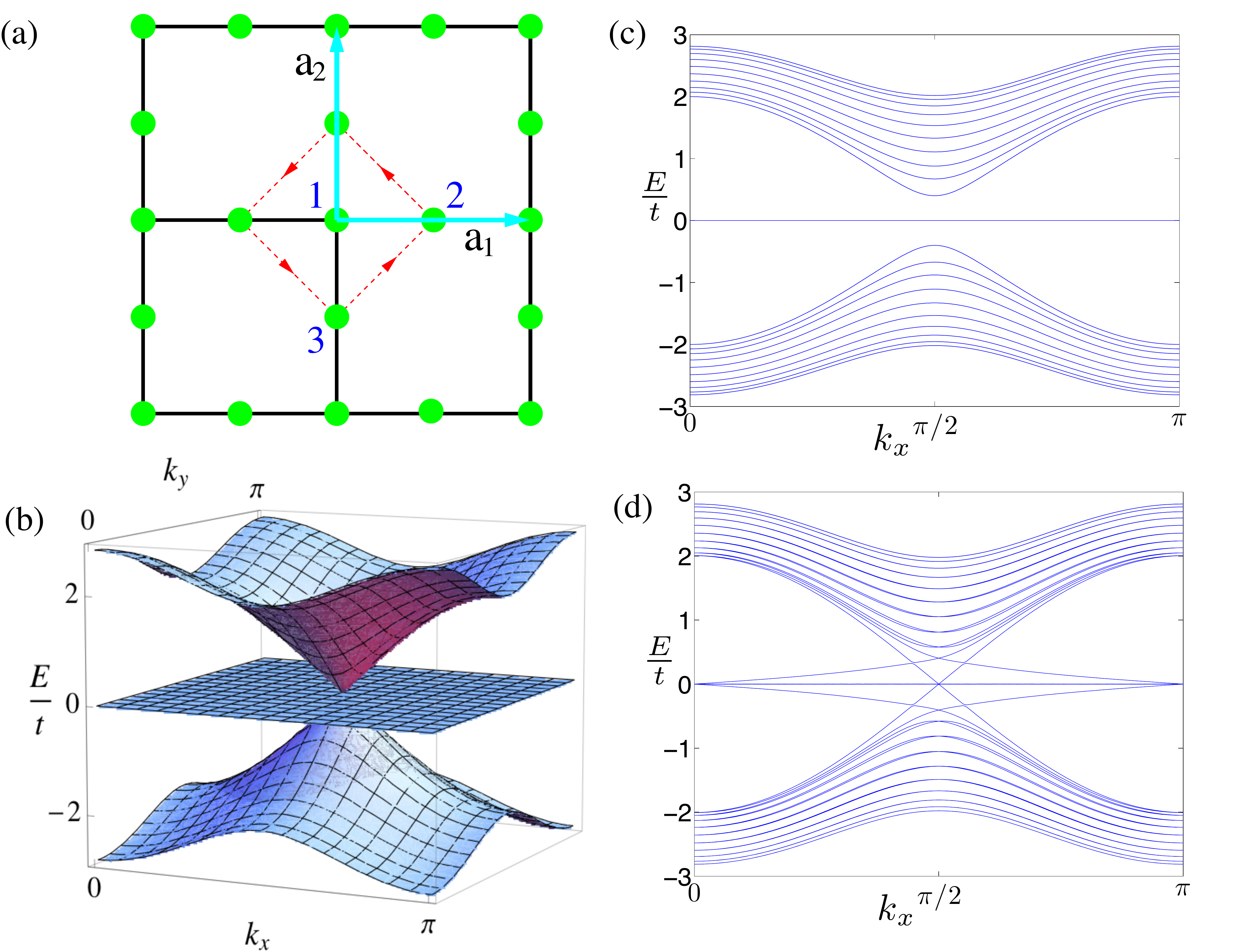}
\caption{ (a) Lieb lattice showing 3-site basis in unit cell, NNN hopping (red dotted lines) and  basis vectors $\mathbf{a}_1$ and $\mathbf{a}_2$. (b) Tight-binding dispersion for ${\cal H}^0_{\bf k}$. (c) Bandstructure for strip of $N_{y}=10$ unit cells with open boundary conditions along the $y$ direction and infinite along $x$ with $\alpha=0.2t$. (d) As in (c), but with $\lambda=0.2t$ instead.} \label{unit_cell}
\end{center}
\end{figure}

We begin with the tight-binding Hamiltonian 

\begin{equation}
\label{h0}
H_{0}=-t\sum_{\langle ij\rangle\alpha}c_{i\alpha}^{\dag}c_{j\alpha} + {\rm h.c},
\end{equation} 
where $c^{\dagger}_{i\alpha}$ creates an electron of spin $\alpha$ on site $i$ of  the so called Lieb lattice, seen in Fig. \ref{unit_cell}a, and $t$ is the hopping amplitude for nearest neighbour sites.

 In momentum space Eq.\
(\ref{h0}) becomes $H_0=\sum_{{\bf k}\sigma}\Psi^\dagger_{{\bf
k}\sigma}{\cal H}^0_{\bf k}\Psi_{{\bf k}\sigma}$ with $\Psi_{{\bf
k}\sigma}=(c_{1{\bf k}\sigma},c_{2{\bf k}\sigma},c_{3{\bf
k}\sigma})^T$ and
\begin{equation}\label{hk0}
{\cal H}^0_{\bf k}=-2t\left(
  \begin{array}{ccc}
    0 & \cos(k_x) & \cos(k_y)  \\
    & 0 &0 \\
    & & 0 \\
  \end{array}
\right),\nonumber
\end{equation}
where the lower triangle of the matrix is understood to be filled so that
the matrix is Hermitian. The spectrum of ${\cal H}^0_{\bf k}$ 
consists of one degenerate flat band $E_{\bf k}^{(3)}=0$ and two
dispersive bands
\begin{equation}\label{ek0}
E_{\bf k}^{(1,2)}=\pm 2t \sqrt{\cos^{2}k_{x}+\cos^{2}k_{y}},
\end{equation}
where the Brillouin zone spans  $-\frac{\pi}{2}\leq k_{x}\leq\frac{\pi}{2}$, $-\frac{\pi}{2}\leq k_{y}\leq\frac{\pi}{2}$.  Plotting these bands yields a low energy Dirac like dispersion (seen in Fig. \ref{unit_cell}b), with a single Dirac cone in the Brillouin zone. At one third filling then, band 1 will be completely filled, with the degenerate flat band and the upper band being empty, and this state
will behave as a gapless band insulator. 

There are several perturbations to the basic Hamiltonian (\ref{h0}) that open up a gap while respecting the translational symmetry of the lattice.  
The simplest option
is to add an on site energy $\epsilon$ to all the sites that have only two nearest neighbours in Fig. \ref{unit_cell}a above, or conversely the inequivalent sites with four nearest neighbours.  We note that this set up has been discussed recently in the context of ultracold fermionic atoms. \cite{Shen} Here, one of the dispersing bands remains touching the flat band while the other dispersing band becomes isolated.  Another option is to add a dimerization term, which staggers the hopping amplitude along both the $\hat{x}$ and $\hat{y}$ directions
so that $t\rightarrow t\pm\alpha$. In this case the flat band becomes isolated from the dispersing bands, which are gapped symmetrically (shown for strip geometry in Fig. \ref{unit_cell}c) . Lastly, including a Rashba spin orbit term will again isolate the flat band, while also splitting the spin degeneracy among the valence and conduction bands.  In all of the above cases the resulting insulating phases are characterized by conventional broken symmetries and are topologically trivial.

The recent paper by Green, Santos and Chamon\cite{chamon} discusses a similar set of energy bands having a flat band directly through the middle of a single Dirac cone, which have been created via staggered flux phases on the kagome lattice. Although their interests lie with the flat band itself, they discuss
possible insulating phases that can be introduced in order to isolate the flat band and determine that, for spinless fermions, time reversal symmetry must be broken in order
to achieve this. Seeing as the Lieb lattice requires no magnetic field to create the same energy band structure, perhaps it is of some interest then that we can isolate the flat band without breaking time reversal symmetry by way of the dimerization term mentioned above. 

Now, to see if this model can support a TI phase  we include the intrinsic SO interaction term
\begin{equation}\label{hso}
H_{\rm SO}=i{\lambda}\sum_{\langle\langle ij
\rangle\rangle\alpha\beta}
 ({\bf d}_{ij}^1\times {\bf d}_{ij}^2)\cdot{\bm \sigma}_{\alpha\beta} c^\dagger_{i\alpha} c_{j\beta},
\end{equation}
where  $\lambda$ is the amplitude for the next-nearest neighbour spin-orbit-induced interaction (shown in Fig. \ref{unit_cell}a), the term $\nu_{ij}%
$=$(\mathbf{{d}}^{1}_{ij}\times\mathbf{{d}}^{2}_{ij})_{z}$=$\pm1$,
where $\mathbf{{d}}^{1}_{ij}$ and
$\mathbf{{d}}^{2}_{ij}$ are the two unit vectors along the nearest neighbour bonds connecting site $i$
to its next-nearest neighbour $j$ and ${\bm \sigma}$ is the vector of Pauli
spin matrices. This will also lead to the formation of a gap at the Dirac point while preserving ${\cal T}$ and the translational symmetry of $H_0$. Fourier transforming, we have
\begin{equation}\label{hkso}
{\cal H}^{\rm SO}_\bk=\pm 4\lambda
\begin{pmatrix}
  0 &0& 0\\
   & 0 & -i\sin{(k_x)} \sin{(k_y)} \\
   &   & 0
\end{pmatrix},
\end{equation}
where the $+(-)$ sign refers to spin up (down) electrons, and the spectrum for the full Hamiltonian, ${\cal H}_\bk={\cal H}^{0}_\bk+{\cal H}^{\rm SO}_\bk$,  then consists of the same degenerate flat band $E_{\bf k}^{(3)}=0$,
and the modified doubly degenerate dispersive bands
\begin{equation}\label{ekSO}
E_{\bf k}^{(1,2)}=\pm 2 \sqrt{t^{2}(\cos^{2}k_{x}+\cos^{2}k_{y})+4\lambda^{2}\sin^{2}k_{x}\sin^{2}k_{y}},
\end{equation}
with a gap  $\Delta_{\rm SO}=4|\lambda|$ at the Dirac point. 

We note that the flat band eliminates the possibility of writing
down a low energy Dirac expansion for the system in terms of the Pauli matrices, as discussed e.g.\ in Ref.\ \onlinecite{Franz1} and other topological insulator models, but, following Ref.\ [\onlinecite{chamon}], it is possible to express the full Hamiltonian above
in terms of a set of $3\times3$ matrices that form a spin-1 representation of $SU(2)$. Although not a focus here, this makes it possible, for instance, to
calculate the Chern numbers of the bands analytically.

To prove that our system is indeed a topological insulator, we now show by an explicit calculation that it possesses a nontrivial  Z$_2$ invariant. There
are several ways to do this in practice, although they are found to be equivalent in the end, so we use the method that is most convenient. 

According to Fu and Kane\cite{Fu2}
when a crystal possesses inversion symmetry the Z$_2$ topological
invariant $\nu$ is
related to the parity eigenvalues $\xi_{2m}({\bm \Gamma}_i)$ of the
2$m$-th occupied energy band at the four ${\cal T}$-invariant momenta
${\bm \Gamma_i}$. Our system is inversion symmetric 
and so we can use this method to find $\nu$. If we select site 1 of the
unit cell as the center of inversion then the parity operator acts as
${\cal P}[\psi_1(\br),\psi_2(\br),\psi_3(\br)]=
[\psi_1(-\br),\psi_2(-\br-\ba_1),\psi_3(-\br-\ba_2)]$ on the triad
of the electron wavefunctions in the unit cell labeled by vector
$\br$. In momentum space the parity operator becomes a diagonal
$3\times 3$ matrix ${\cal P}_\bk= {\rm diag}(1,e^{-i\ba_1\cdot
\bk},e^{-i\ba_2\cdot \bk})$ and the four ${\cal T}$-invariant momenta
can be expressed as
${\bm \Gamma}_i=\pi(\hat{x}n_i+\hat{y}m_i)/2$
with $n_i,m_i=0,1$.  The eigenstates
of ${\cal H}_{{\bm \Gamma}_i}$ can be found analytically in this case
and it is then easy to determine the parity
eigenvalues of the occupied bands. At ${1\over 3}$ filling we find that three $\xi$'s are
positive and one is negative.  Which of the four $\xi$'s is negative
depends on the choice of the inversion center but the product
$\Pi_i\xi({\bm \Gamma}_i)=(-1)^\nu$ is independent of this choice and
determines the non-trivial Z$_2$ invariant $\nu=1$, confirming that the
system is indeed a topological insulator. Similar considerations for
${2\over 3}$ filling also yield $\nu=1$. 

Next, we solve the model  in a strip geometry numerically using exact diagonalization, in order to incorporate edge effects and demonstrate the bulk-boundary correspondence, which states that whenever $\nu=1$ there will be a pair of topologically protected gapless modes along each edge in the system. For the trivial insulating state with $\alpha=0.1t$, we find no edge states (Fig. \ref{unit_cell}c), but for $\lambda=0.1t$, we indeed find a pair of spin-filtered gapless states associated with each edge, as seen in Fig. \ref{unit_cell}d. As there are 3 atoms in the unit cell, the two edges of the slab are not equivalent, and therefore
the edge states are not degenerate.
 
\section{perovskite Lattice}
\begin{figure}[t]
\begin{center}
\includegraphics[scale=0.192]{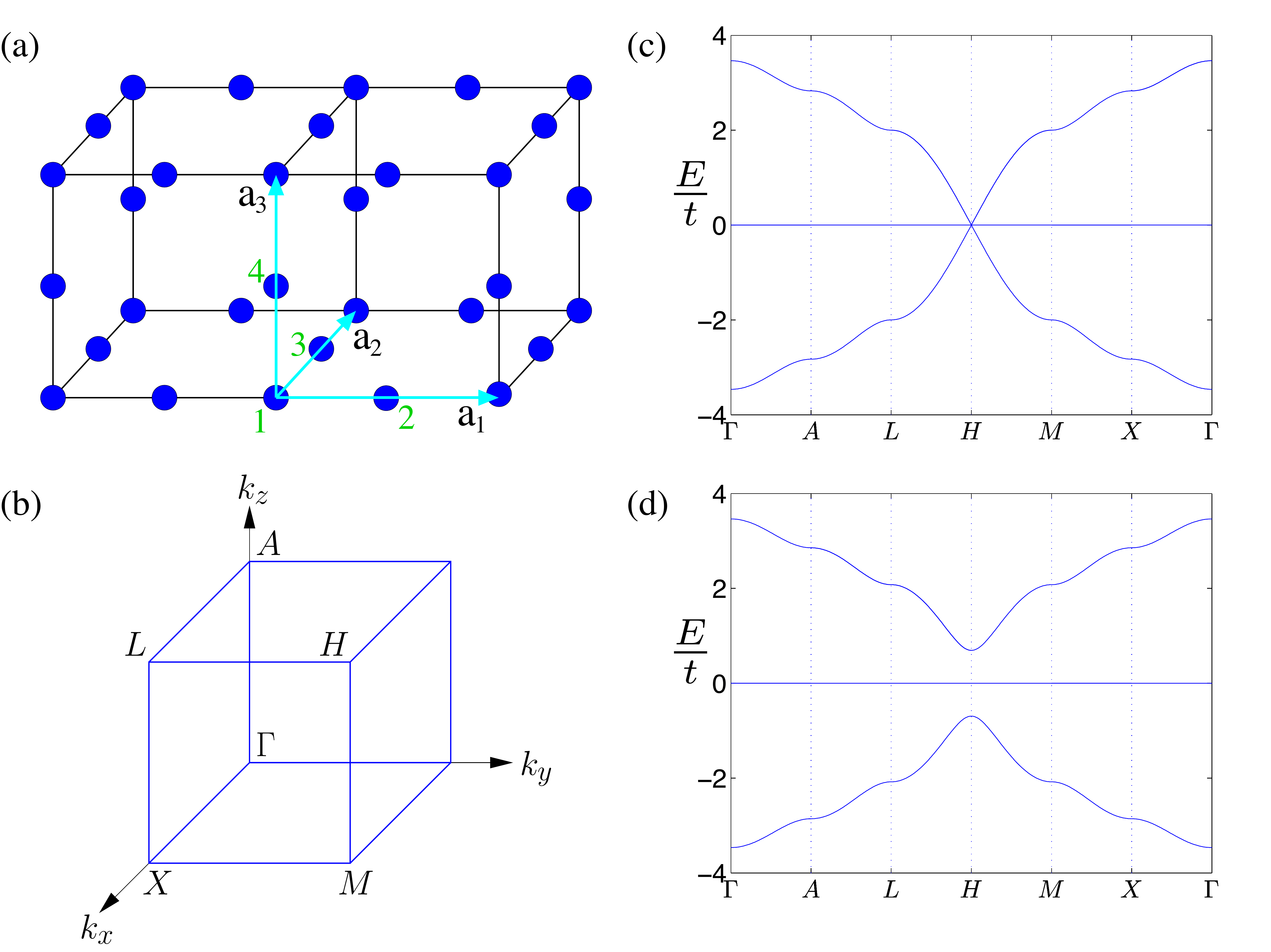}
\caption{ (a) Perovskite lattice showing 4 sites in unit cell along with basis vectors. (b) High symmetry points in the Brillouin zone. (c) Bandstructure inside bulk along path of high symmetry for ${\cal H}^0_{\bf k}$. (d) As in (c), but with $\alpha=0.2t$. } \label{perovskite}
\end{center}
\end{figure}

The perovskite lattice, depicted in Fig. \ref{perovskite}a, is a straightforward generalization of the Lieb lattice into 3D and can be viewed as a simple cubic lattice with additional sites positioned at the centers of all edges. 

As with the Lieb lattice, our starting point is the tight binding  model given in Eq. (\ref{h0}), which in momentum space gives 
\begin{equation}\label{hk0p}
{\cal H}^0_{\bf k}=-2t\left(
  \begin{array}{cccc}
    0 & \cos{k_x} & \cos{k_y} & \cos{k_z} \\
    & 0 &0& 0 \\
    & & 0 & 0\\
    & & &0 \\
  \end{array}
\right).\nonumber
\end{equation}
The spectrum then consists of two degenerate flat bands $E_{\bf k}^{(3,4)}=0$ and the two dispersive bands
\begin{equation}\label{ek0}
E_{\bf k}^{(1,2)}=\pm2t\sqrt{\cos^{2}k_{x}+\cos^{2}k_{y}+\cos^{2}k_{z}}.
\end{equation}
Again, we have a single Dirac point in the Brillouin zone and at ${1\over 4}$ filling the system behaves as a gapless band insulator. 
 The spectrum is shown in Fig. \ref{perovskite}c.  As above, we can create a topologically trivial insulating phases by including an anisotropic on site term or a dimerization (Fig. \ref{perovskite}d). 
 
\begin{figure}[t]
\begin{center}
\includegraphics[scale=0.295]{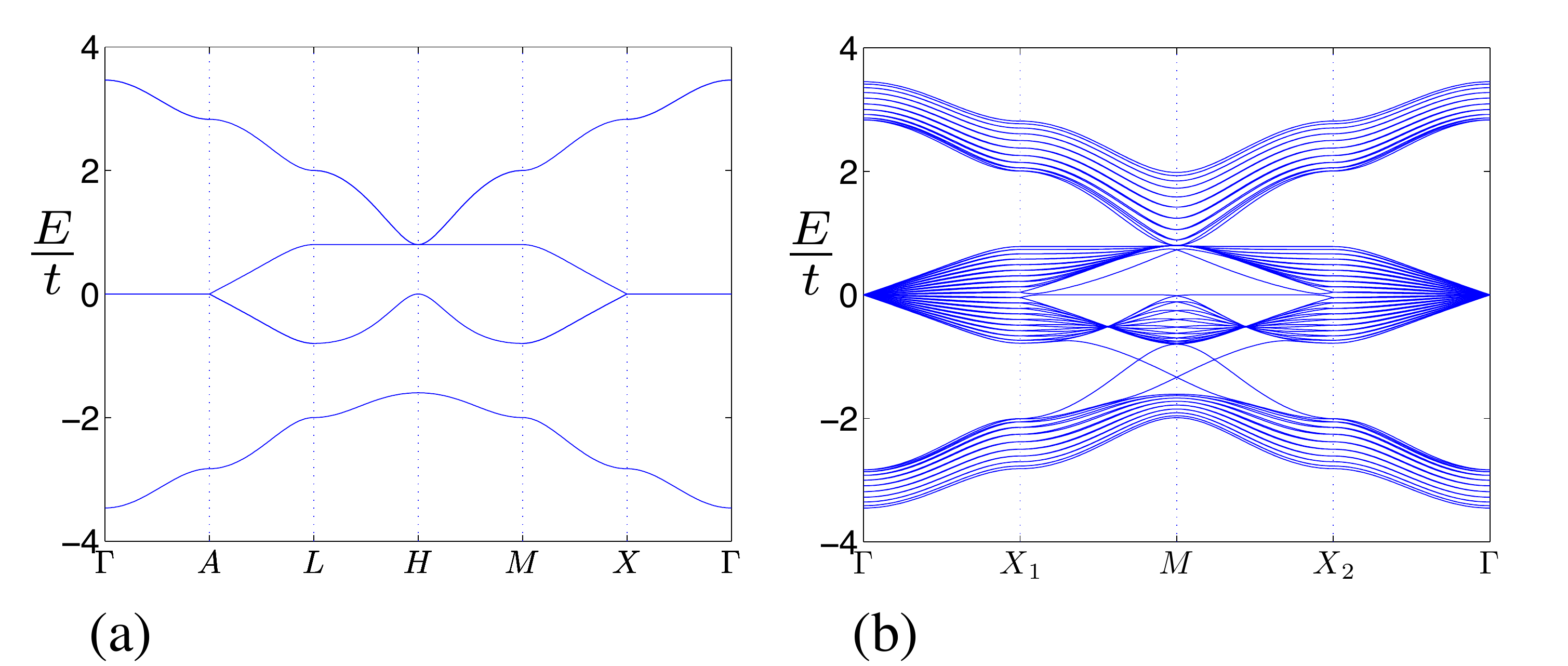}
\caption{ (a) Bulk bandstructure shown along path connecting high symmetry points for $\lambda=0.2t$.  (b) Bandstructure for slab geometry along lines connecting four surface TRIM for $\lambda=0.2t$.  } \label{perovskiteII}
\end{center}
\end{figure}

 To open up a topologically non-trivial gap, we consider spin-orbit coupling Eq. (\ref{hso}). Since the ${\bf d}_{ij}^{1,2}$ now lie in three-dimensional
space, the Hamiltonian does not decouple for the two spin projections as above, and instead
becomes an $8\times 8$ matrix in $k$-space. The spectrum can be found numerically and is shown in Fig. \ref{perovskiteII}a. Unlike the model on the Lieb lattice, the addition of the spin orbit term creates a dispersion in the degenerate flat bands and a gap that depends on the sign of $\lambda$, which is no longer symmetric around zero energy.  Switching the sign of $\lambda$ inverts the band structure.

We now study the topological classes of these insulating phases. As 
above, the $Z_2$ topological invariants $(\nu_0;
\nu_1 \nu_2 \nu_3)$ are easy to evaluate when a crystal possesses
inversion symmetry and can be determined
from knowledge of the parity eigenvalues $\xi_{2m}({\bm \Gamma}_i)$ of
the 2$m$-th occupied energy band at the 8 ${\cal T}$-invariant momenta
(TRIM) $\BGamma_i$ that satisfy $\BGamma_i = \BGamma_i + {\bf G}$. The 8
TRIM in our system can be expressed in terms of primitive
reciprocal lattice vectors as $\BGamma_{i=(n_1 n_2 n_3)} = (n_1 {\bf
b}_1 + n_2 {\bf b}_2 + n_3 {\bf b}_3)/2$, with $n_j = 0,1$. Then
$\nu_\alpha$ is determined by the product $(-1)^{\nu_0} = \prod_{n_j =
0,1} \delta_{n_1 n_2 n_3},$ and $(-1)^{\nu_{i=1,2,3}} = \prod_{n_{j\ne i}
= 0,1; n_i = 1} \delta_{n_1 n_2 n_3}$, where
$\delta_i=\prod_{m=1}^{N} \xi_{2m}(\BGamma_i)$.

If we select site 1 of the unit cell, Fig.\
\ref{perovskite}, as the center of inversion then the parity operator
acts as ${\cal P}[\psi_1({\bf r}),\psi_2({\bf r}),\psi_3({\bf
r}),\psi_4({\bf r})]= [\psi_1(-{\bf r}),\psi_2(-{\bf
r}-{\bf a}_1),\psi_3(-{\bf r}-{\bf a}_2),\psi_4(-{\bf r}-{\bf a}_3)]$ on the
four-component electron wave function in the unit cell labeled by
vector ${\bf r}$. In momentum space and including spin the parity
operator becomes a diagonal $8\times 8$ matrix ${\cal P}_{\bf k}=
{\rm diag}(1,e^{-i{\bf a}_1\cdot {\bf k}},e^{-i{\bf a}_2\cdot {\bf
k}},e^{-i{\bf a}_3\cdot {\bf k}})\otimes {\rm diag}(1,1)$. It is
straightforward to obtain the eigenstates of ${\cal H}_{{
\BGamma}_i}$ and the parity eigenvalues of the occupied bands
numerically, then determine the $Z_2$ invariants.
At quarter filling,
we find that $\delta=-1$ at the $H$ point and $\delta=1$ at the other
TRIM, so the spin-orbit phase is a $(1;111)$ strong topological
insulator.

Incorporating edge effects into the system with a slab geometry and plotting the band energies along lines connecting the four surface TRIM (Fig. \ref{perovskiteII}b), we can also see the bulk energy bands and an odd number of surface states which traverse the gap, a behavior characteristic of STI.

\section{Conclusions}
We have established a model for a topological insulator on lattices with simple cubic symmetry.  In 2D we have shown that a tight-binding model with spin-orbit coupling on the Lieb lattice supports gapless topologically protected edge modes, and possesses a non-trivial $Z_{2}$ invariant.  In 3D, we have demonstrated that a similar model on the perovskite lattice supports  2D gapless surface states when in the (1;111) STI phase, brought on by spin orbit coupling. We have also identified other gapped phases in these models that are topologically trivial.

In addition to the topologically non-trivial band structures, the models considered here exhibit flat bands, which could give rise to interesting strongly correlated electron states even for relatively weak interaction strength.\cite{balents1} A necessary condition for non-trivial correlated physics is that  the single-electron states forming the flat band are delocalized in space (localized states typically lead to formation of a Wigner crystal). Delocalized single-electron states are guaranteed to exist if the flat band possesses a non-zero topological invariant, as happens e.g.\ in quantum Hall liquids. Unfortunately, this is not the case in models considered in this study. In the Lieb lattice spin-orbit coupling preserves the flat band and separates from other bands. The resulting flat band however turns out to be topologically trivial, i.e. it has $\nu=0$, while the TI behavior derives from the two dispersive bands which have $\nu=1$.  In the perovskite lattice spin-orbit coupling produces a significant dispersion in the original flat band and this will limit the importance of correlations in the system. 

Can our model be realized in a physical system?
There exist many perovskites in nature as well as many layered perovskites composed of weakly coupled 2D planes with Lieb lattice structure. The most prominent example of the latter are the CuO$_2$ planes in high-$T_c$ cuprate superconductors such as YBa$_2$Cu$_3$O$_7$ or Bi$_2$Sr$_2$CaCu$_2$O$_8$.  In 3D, our model system is an idealization of the naturally occurring perovskites, most of which also
have a heavy central atom in the middle of each cubic cell, such as SrTiO$_{3}$ or the double perovskite structure Ba$_{2}$NaOsO$_{6}$. In these real materials the electron behavior near the Fermi level derives from the $e_g$ and $t_{2g}$ orbitals of the transition metal element occupying the cubic site while the edge sites are normally oxygens whose $p$-orbitals are far away from the Fermi level. The resulting band structure for the active orbitals is then significantly more complex than that captured in our simple tight binding model. Nevertheless, our model calculations demonstrate that this class of tight-binding Hamiltonians on lattices with cubic symmetry can support topological phases, both in 2D and 3D. We hope that our work will stimulate detailed band structure calculations of perovskites with heavy transition metal elements in search for new families of topological insulators.

In the more exotic realm, it might be possible to artificially engineer the 2D system by modulating a two-dimensional electron gas with a periodic potential having Lieb symmetry, as achieved recently in constructing the `artificial graphene'.\cite{west1} Another possibility lies with cold Fermionic atoms in optical lattices as discussed in Refs.\ [\onlinecite{DasSarma, Zhang3}].

\emph{Acknowledgment.}---The authors have benefited from discussions with H.-M. Guo and C. Felser. This work was supported by NSERC and CIfAR .


\end{document}